\documentclass[12pt]{revtex4-1}
\usepackage{amsfonts,amssymb,amsmath,graphicx}
\begin{document}

\title{Two-point correlation function for Dirichlet $L$-functions}
\author{E. Bogomolny$^{1,2}$, J.P. Keating$^3$}
\affiliation{$^{1}$Univ. Paris-Sud,
Laboratoire de Physique Th\'eorique et Mod\`eles Statistiques,
Orsay, F-91405, France,\\
$^2$CNRS, UMR8626, Orsay, F-91405, France\\
$^{3}$ School of Mathematics,\\
University of Bristol, Bristol, BS8 1TW, UK}

\begin{abstract}
The two-point correlation function for the zeros of Dirichlet $L$-functions at a height $E$ on the critical line is calculated heuristically using a generalization of the Hardy-Littlewood conjecture for pairs of primes in arithmetic progression. The result matches the conjectured Random-Matrix form in the limit as $E\rightarrow\infty$ and, importantly, includes finite-$E$ corrections.  These finite-$E$ corrections differ from those in the case of the Riemann zeta-function, obtained in (1996 \textit{Phys. Rev. Lett.} \textbf{77} 1472), by certain finite products of primes which divide the modulus of the primitive character used to construct the $L$-function in question.      
\end{abstract}

\maketitle
\section{Introduction}

The observation that statistical properties of the zeros of the Riemann zeta function are the same as those of the eigenvalues of unitary ensembles of random matrices has a long history \cite{Mont, Odlyzko} (see \cite{KS} for a review). The main difficulty in proving this is the calculation of off-diagonal prime contributions to correlation functions when these are represented using the explicit formula that relates the zeros to the primes.  It appears that to calculate these contributions it is necessary to use the Hardy-Littlewood conjecture \cite{HL} for pairs of prime numbers.  In  \cite{keating_2} a smooth version of this conjecture was used to demonstrate that the two-point correlation function of the Riemann zeros in the universal limit coincides with that of the Unitary Ensembles (e.g.~the Gaussian (GUE), or Circular (CUE) Unitary Ensembles) of Random Matrix Theory (RMT). In \cite{BK}, using the same smoothed form of the Hardy-Littlewood conjecture, it was shown that all correlation functions of the Riemann zeros agree with the corresponding GUE/CUE results.    

A more precise expression for the two-point correlation function of  the Riemann zeta function zeros was obtained in \cite{keating} using another method which is equivalent to the full Hardy-Littlewood conjecture (details of the calculations can be found in \cite{varenna}).  The conjectured expression for the two-point correlation function reads
\begin{equation}
R_2(\epsilon)=\bar{d}^2(E)+R_2^{\,\mathrm{(diag)}}(\epsilon)+R_2^{\,\mathrm{(off)}}(\epsilon).
\label{R_riemann}
\end{equation}
Here, if $d(E)$ denotes the density of zeros at height $E$ on the critical line and
\begin{equation}
\bar{d}(E)=\frac{1}{2\pi}\ln \left (\frac{E}{2\pi} \right )
\end{equation}
its asymptotic mean (defined to be the derivative of the non-fluctuating part of the counting function of the zeros, 
$N(E)=\frac{E}{2\pi}\ln\left (\frac{E}{2\pi}\right )-\frac{E}{2\pi} +O(\ln E)$), then
\begin{equation}
R_2(\epsilon)=\left< d(E-\epsilon/2)d(E+\epsilon/2)\right>.
\end{equation}

The diagonal part of the correlation function, $R_2^{\,\mathrm{(diag)}}(\epsilon)$, has the form
\begin{equation}
R_2^{\mathrm{\,(diag)}}(\epsilon)=
-\frac{1}{4\pi^2}\frac{\partial^2}{\partial \epsilon^2}\ln \left [|\zeta(1+\mathrm{i}\epsilon)|^2\Phi^{\mathrm{\,(diag)}}(\epsilon)\right ]
\label{R_diag_riemann}
\end{equation}
where 
\begin{equation}
\Phi^{\mathrm{\,(diag)}}(\epsilon)=\prod_p \exp \left ( 2\sum_{m=1}^{\infty}\frac{1-m}{m^2p^m}\cos(m\epsilon \ln p) \right  ).
\label{psi_diag_riemann}
\end{equation}
The off-diagonal part of the correlation function, $R_2^{\,\mathrm{(off)}}(\epsilon)$ is 
\begin{equation}
R_2^{\,\mathrm{(off)}}(\epsilon)= 
\frac{1}{4\pi^2 } \mathrm{e}^{2\pi \mathrm{i}\epsilon \bar{d}(E)} |\zeta(1+\mathrm{i}\epsilon)|^2
\Phi^{\,(\mathrm{off})}(\epsilon) 
+\mathrm{c.c.}
\label{R_off_riemann}
\end{equation}
where 
\begin{equation}
\Phi^{\,(\mathrm{off})}(\epsilon)=\prod_{p}\left (1-\frac{(p^{\mathrm{i}\epsilon}-1)^2}{(p-1)^2} \right ). 
\label{psi_off_riemann}
\end{equation}
The assumed (optimistic) precision of  these formulas is $\mathcal{O}(E^{-1/2})$ and they agree very well with numerical calculations of Odlyzko \cite{SIAM, osaka}.  In the limit $E\rightarrow\infty$, $R_2(\epsilon/\bar{d}(E))/(\bar{d}(E))^2$ tends to the GUE/CUE expression for the two-point correlation function.

In \cite{BL} the method of \cite{keating_2} was generalized to show that the two-point correlation function of the zeros of a given Dirichlet $L$-function tends, in the universal limit (i.e.~infinitely high up the critical line), to the GUE/CUE result. 

The purpose of this note is to obtain an expression, similar to the above formulas, for the two-point correlation function of Dirichlet $L$-functions  using the full Hardy-Littlewood conjecture. The final answer is similar to the Riemann case (\ref{R_diag_riemann})-(\ref{psi_off_riemann}) but with a finite contribution from primes which divide the modulus of the character defining the $L$-function in question.  This demonstrates that the method of \cite{keating} applies in much greater generality than to the zeta function alone: it extends to other $L$-functions.  The plan of the note is as follows. In Sec.~\ref{mean}  the appropriate form of the  conjecture for  prime pairs in arithmetical progressions is obtained. In Sec.~\ref{trace} the explicit formula for Dirichlet $L$-functions is stated and in Sec.~\ref{two_points} the formal expression for the two-point correlation function of their zeros is presented. In Sec.~\ref{diagonal} the diagonal contribution is derived and in Sec.~\ref{off_diagonal} the prime pairs conjecture is used to derive the off-diagonal contribution.  The list of final formulas is given in Sec.~\ref{conclusion}. In Appendix~\ref{zero_term} it is checked that the GUE/CUE result corresponds to the contribution of only one term, as shown in \cite{BL} by a different method.

For the Riemann zeta-function, the expression for the two-point correlation function, (\ref{R_diag_riemann})-(\ref{psi_off_riemann}), that includes the lower order terms may also be obtained by a heuristic recipe \cite{CFKRS, CFZ} for calculating averages of products and ratios of zeta-functions \cite{CS}.  That approach extends to $L$-functions too.  It has the advantage of being simple and powerful in its generality, but the disadvantage that some of the key steps can only be justified under the assumption that certain large error terms must cancel each other, for reasons not yet understood.  Our goal in this work is to extend the explicit approach based on the Hardy-Littlewood conjecture to the more general case of $L$-functions. 

\section{Prime pairs in arithmetical progression}\label{mean}

Let  $\Lambda(m)$ denote the von Mangoldt function: $\Lambda(m)=\log p$ if $m$ is a power of a prime $p$ and $\Lambda(m)=0$ otherwise.  The usual Hardy-Littlewood conjecture for the density of prime pairs can be written in two equivalent forms \cite{HL}: either
\begin{equation}
\lim_{N\to \infty} \frac{1}{N}\sum_{m=1}^N \Lambda(m) \Lambda(m+h)=\alpha(h),
\end{equation} 
or, if $\pi_2(h; N)$ denotes the number of primes $p<N$ such that $p+h$ is also a prime, then
\begin{equation}
\pi_2(h; N)\overset{N\to\infty}{\longrightarrow} \alpha(h)\frac{N}{\ln^2N}.
\end{equation}
Here $\alpha(h)$ is defined as follows: $\alpha (h)=0$ if $h$ is odd, and if $h$ is even
\begin{equation}
\alpha (h)=\sum_{q=1}^{\infty} \left ( \frac{\mu(q)}{\phi(q)} \right )^2 c_q(h)=2C_2\prod_{p\,|\,h}\frac{p-1}{p-2},
\label{alpha_h}
\end{equation}
where $c_q(n)$ are the Ramanujan coefficients
\begin{equation}
c_q(n)=\sum_{(p,q)=1} \mathrm{e}^{2\pi\mathrm{i}pn/q},
\end{equation}
$\mu(q)$ is the M\"obius function, $\phi(q)$ is Euler's totient function, and $C_2$ is the twin prime constant
\begin{equation}
C_2=\prod_{p>2}\left (1-\frac{1}{(p-1)^2}\right )\approx 0.6601618\; .
\end{equation}
To calculate the off-diagonal part of the two-point correlation function of the zeros of a Dirichlet $L$-function we need to know the mean value of the product of two $\Lambda$-functions in arithmetic progression
\begin{equation}
\alpha(r_2-r_1,k)=\lim_{N\to \infty} \frac{1}{N}\sum_{m=1}^N \Lambda(km+r_1) \Lambda(km+r_2). 
\end{equation}
It is plain that in order that progressions $mk+r_i$ contain an infinite number of primes it is necessary that $(k,r_1)=1$ and $(k,r_2)=1$. In the above definition we anticipate that under these conditions the indicated limit depends only on the difference between $r_2$ and $r_1$.
  
This limit can heuristically be obtained from usual probabilistic considerations \cite{keating_2}, but we found it more convenient to  
start with the exact expansion of the $\Lambda$-function as a series of the Ramanujan coefficients \cite{Hardy}
\begin{equation}
\frac{\phi(n)\Lambda(n)}{n}=\sum_{q=1}^{\infty}\frac{\mu(q)}{\phi(q)}c_q(n).
\label{hardy}
\end{equation}
Using the following properties of the Ramanujan coefficients
\begin{eqnarray}
&\mathrm{if}\;&m\equiv n\;\mathrm{mod}\;q,\;\;\;c_q(m)=c_q(n),\\
&\mathrm{if}\;&(q,r)=1,\;\;\;c_{qr}(n)=c_q(n)c_r(n),
\end{eqnarray}
one can easily prove that 
for squarefree integers $q_1$ and $q_2$ (as we will multiply them by $(\mu(q_1) \mu(q_2))^2$) one has
\begin{equation}
\lim_{N\to \infty} \frac{1}{N}\sum_{m=1}^N c_{q_1}(km+r_1)c_{q_2}(km+r_2)=c_v(r_1-r_2)c_{\tilde{q}_1}(r_1)c_{\tilde{q}_2}(r_2)
\end{equation}
where $q_i=v\tilde{q}_i$ and $(v,k)=1$ but $\tilde{q}_1\,|k$ and $\tilde{q}_2\,|k$. In all other cases the above mean value is zero.

Using (\ref{hardy}) and formally interchanging the summation and the average (which is not justified rigorously) after simple calculations we get
\begin{equation}
\alpha(r_2-r_1,k)= S(k)
\sum_{(q,k)=1}\left (\frac{\mu(q)}{\phi(q)}\right )^2c_q(r_2-r_1)
\label{HL}
\end{equation}
with
\begin{eqnarray}
S(k)&=&\sum_{v_1\,|\,k}\frac{\mu(v_1)}{\phi(v_1)}c_{v_1}(r_1)\sum_{v_2\,|\,k}\frac{\mu(v_2)}{\phi(v_2)}c_{v_1}(r_2)
\nonumber\\
&=& \prod_{p\, |\, k}\left ( 1-\frac{c_p(r_1)}{\phi(p)}\right )\prod_{p\, |\,k}\left ( 1-\frac{c_p(r_2)}{\phi(p)})\right )\nonumber\\
&=&\left \{ \begin{array}{cc}\prod_{p\,|\,k}\left (\frac{p}{p-1} \right )^{2}=\left (\frac{k}{\phi(k)}\right )^2,&(r_1,k)=1,\;(r_2,k)=1\\0,&\;\mathrm{otherwise}\end{array}\right . .
\label{S_k}
\end{eqnarray}
In the last step we used that for prime $p$
\begin{equation}
c_p(n)=\left\{ \begin{array}{cl}-1,\;&\mathrm{if}\;p \nmid n\\ \phi(p),\;&\mathrm{if}\;p\,|n\end{array}\right . .
\end{equation}
The expression (\ref{S_k}) implies that the Hardy-Littlewood conjecture for prime pairs in arithmetical progression can be formulated as follows. Consider an arithmetical progression $km+r$ with $(k,r)=1$, the number of primes in this progression with $m<N$ such that $p+h$  is also a prime ($(k,r+h)=1$) is asymptotically
\begin{equation}
(\#\  p,p+h\; \mathrm{primes},\;p\in mk+r,\;m<N)\overset{N\to\infty}{\longrightarrow}\alpha(h,k)\frac{N}{\ln^2N}. 
\end{equation}  
The expression for $\alpha(h,k)$ as a product of prime numbers has the form
\begin{equation}
\alpha(h,k)=\alpha(h)\beta(h,k)
\end{equation}
where $\alpha(h)$ is the same as in the usual Hardy-Littlewood conjecture (\ref{alpha_h}) and $\beta(h,k)$ is a finite product of primes which divide $k$
\begin{equation}
\beta(h,k)=\prod_{\underset{\scriptstyle{p\,|\,h}}{p\,|\,k}}\frac{p-1}{p-2}\prod_{p\,|\,k}\frac{(p-1)^2}{p(p-2)}.
\end{equation}

\section{Explicit formula }\label{trace}

Dirichlet $L$-functions are defined as follows
\begin{equation}
L(s,\chi)=\sum_{n=1}^{\infty}\frac{\chi(n)}{n^s}=\prod_{p}\left (1-\frac{\chi(p)}{p^s} \right )^{-1}
\end{equation}
where $\chi(n)$ is a character modulo an integer $k$. Below we shall assume that the character is primitive. 
The Generalized Riemann Hypothesis (GRH) states that all non-trivial zeros of this $L$-function lie on the critical line $s=1/2+\mathrm{i}E$ with real $E$; that is, the zeros may be written $s=1/2+\mathrm{i}E_n$ where $E_n$ is real.  We are interested in the statistical properties of these zeros as $E\to \infty$.
 
The explicit formula for the density of zeros of any Dirichlet $L$-function  has the form
\begin{equation}
d(E)=\sum_n\delta(E-E_n)=\bar{d}(E)+d^{\mathrm{osc}}(E)
\end{equation}
where the smooth part is, asymptotically,
\begin{equation}
\bar{d}(E)=\frac{1}{2\pi}\ln \left ( \frac{kE}{2\pi}\right )
\label{d_dirichlet}
\end{equation}
and the oscillating part is
\begin{equation}
d^{\,\mathrm{(osc)}}(E)=-\frac{1}{2\pi}\sum_n\frac{\Lambda(n)}{\sqrt{n}}\chi(n)\mathrm{e}^{\mathrm{i}E\ln n}+\mathrm{c.c.} 
\end{equation}

\section{Two-point correlation function}\label{two_points}

The formal expression for the two-point correlation function is
\begin{equation}
R_2(\epsilon_1,\epsilon_2)=\langle d(E+\epsilon_1)d(E+\epsilon_2)\rangle\approx
\bar{d}^2(E)+R_2^{\,\mathrm{(osc)}}(\epsilon_1,\epsilon_2)
\end{equation} 
with
\begin{eqnarray}
R_2^{\,\mathrm{(osc)}}(\epsilon_1,\epsilon_2)&=&\langle d^{\,\mathrm{(osc)}}(E+\epsilon_1)d^{\,\mathrm{(osc)}}(E+\epsilon_2)  \rangle \nonumber\\
&=&\frac{1}{4\pi^2}\langle \sum_{n_1,n_2}\frac{\Lambda(n_1)\Lambda(n_2)}{\sqrt{n_1n_2}}\chi(n_1)\bar{\chi}(n_2)
\mathrm{e}^{\mathrm{i}(E+\epsilon_1)\ln n_1-\mathrm{i}(E+\epsilon_2)\ln n_2}\rangle +\mathrm{c.c.}\\
&=& R_2^{\mathrm{\,(diag)}}(\epsilon_1,\epsilon_2)+R_2^{\,\mathrm{(off)}}(\epsilon_1,\epsilon_2).
\nonumber
\end{eqnarray}

\subsection{Diagonal contribution}\label{diagonal}

The diagonal part of the two-point function consists of taking into account only the terms with $n_1=n_2=n$
\begin{equation}
R_2^{\mathrm{\,(diag)}}(\epsilon_1,\epsilon_2)=
\frac{1}{4\pi^2}\sum_{n}\frac{\Lambda^2(n)}{n}|\chi(n)|^2
\mathrm{e}^{\mathrm{i}(\epsilon_1-\epsilon_2)\ln n} +\mathrm{c.c.}\; .
\end{equation}
For any character $|\chi(n)|^2=1$ for $(n,k)=1$ and $|\chi(n)|^2=0$ for $(n,k)>1$. Therefore, setting $\epsilon=\epsilon_1-\epsilon_2$,
\begin{equation}
R_2^{\mathrm{\,(diag)}}(\epsilon)=
\frac{1}{4\pi^2}\sum_{(n,k)=1}\frac{\Lambda^2(n)}{n}
\mathrm{e}^{\mathrm{i}\epsilon\ln n} +\mathrm{c.c.}\; .
\end{equation}
The calculation of this expression can be carried out exactly as for the Riemann case, but removing terms that divide $k$: 
\begin{equation}
R_2^{\mathrm{\,(diag)}}(\epsilon)=
-\frac{1}{4\pi^2}\frac{\partial^2}{\partial \epsilon^2}\ln \left [|\zeta(1+\mathrm{i}\epsilon)|^2\Phi^{\mathrm{\,(diag)}}(\epsilon) \Psi(\epsilon,k) \right ]
\end{equation}
where $\Phi^{\mathrm{\,(diag)}}(\epsilon)$ is the same function as in the Riemann case (\ref{psi_diag_riemann}) 
and
\begin{equation}
\Psi(\epsilon,k)=\exp \left (-\sum_{p| k}\sum_{m=1}^{\infty}\frac{1}{m^2 p^m}\mathrm{e}^{\mathrm{i}m\epsilon \ln p}+\mathrm{c.c.}\right ).
\label{psi_diag_dirichlet}
\end{equation}

\subsection{Off-diagonal contribution}\label{off_diagonal}

The off-diagonal part of the correlation function corresponds to $n_1=n$, $n_2=n+h$ with $h\ll n$
\begin{equation}
R_2^{\,\mathrm{(off)}}(\epsilon)\approx 
\frac{1}{4\pi^2}\langle\  \sum_{n,h}\frac{\Lambda(n)\Lambda(n+h)}{n}\chi(n)\bar{\chi}(n+h)
\mathrm{e}^{\mathrm{i}\epsilon\ln n-\mathrm{i}E h/n }\ \rangle +\mathrm{c.c.}\; .
\end{equation}
Now choose $n=km+r_1$ and $h=kl+r$ with $r_1$ and $r$ modulo $k$ (but, of course, $r_1\neq 0$ and $r+r_1\neq 0$ mod $k$) 
\begin{eqnarray}
R_2^{\,\mathrm{(off)}}(\epsilon)&\approx & 
\frac{1}{4\pi^2}  \sum_{r_1,r=0}^{k-1}\chi(r_1)\bar{\chi}(r_1+r)\nonumber\\ 
&\times &\sum_{m,l}\frac{\Lambda(km+r_1)\Lambda(km+r_1+kl+r)}{km}
\mathrm{e}^{\mathrm{i}\epsilon\ln km-\mathrm{i}E(kl+r)/(km)} +\mathrm{c.c.}\; .
\end{eqnarray}
Using (\ref{HL}) for the mean value of the product of the two $\Lambda$-functions one gets
\begin{eqnarray}
R_2^{\,\mathrm{(off)}}(\epsilon)&\approx & 
\frac{1}{4\pi^2}S(k)  \sum_{r_1,r=0}^{k-1}\chi(r_1)\bar{\chi}(r_1+r) \nonumber\\
&\times &\sum_{(q,k)=1}\sum_{(p,q)=1}\left (\frac{\mu(q)}{\phi(q)}\right )^2
\sum_{m,l}\frac{1}{km}
\mathrm{e}^{\mathrm{i}\epsilon\ln km-\mathrm{i}E(kl+r)/(km)+2\pi \mathrm{i}p(kl+r)/q} +\mathrm{c.c.}\; .
\end{eqnarray}
As in the Riemann case, the summation over $l$ and $m$ can be performed as follows. The sum over all $l$ gives, using the Poisson Summation Formula,
\begin{equation}
\sum_{l=-\infty}^{\infty}\mathrm{e}^{2\pi \mathrm{i}(kl+r)x}=\sum_{s=-\infty}^{\infty}
 \delta(k x-s) \mathrm{e}^{2\pi \mathrm{i}xr}=\sum_{s=-\infty}^{\infty}
 \delta(k x-s)\mathrm{e}^{2\pi \mathrm{i}sr/k}
\end{equation}
with 
\begin{equation}
x=\frac{p}{q}-\frac{E}{2\pi km}
\end{equation}
In the above formulas $p$ is coprime to $q$ and smaller than $q$, i.e. $0<p<q$. To take into account the periodicity of the $\delta$-function above it is convenient to define the integer
\begin{equation}
R=kp-qs.
\label{normal_form}
\end{equation}
It is coprime to $q$ because $p$ and $k$ are coprime to $q$. 

By construction $(k,q)=1$. Therefore there exist two integers $a$ and $b$ such that $(a,q)=1$, $(b,k)=1$ and 
\begin{equation}
ak-bq=1.
\end{equation}
Any number $M$ can be written as $M=pk-qs$. Integers $p$ and $s$ are not unique but have the form $p=aM+tq$ and  $s=bM+tk$. If we require that $(M,q)=1$ then uniquely $aM=r_m+q t_m$ with $(r_m,q)=1$ and $0<r_m<q$. Now we get uniquely $p=r_m$ and $t=-t_m$. This means that any integer $R$ coprime with $q$ can be uniquely rewritten in the form 
(\ref{normal_form}) with $0<p<q$, and the summation over $(p,q)=1$ with $0<p<q$ and all $s$ is equivalent to the summation over $(R,q)=1$ without restriction on $R$. 

One has
\begin{eqnarray}
& &\sum_{m,l}\frac{1}{m}
\mathrm{e}^{\mathrm{i}\epsilon\ln km-\mathrm{i}E(kl+r)/m+2\pi \mathrm{i}p (k l+r)/q}\nonumber \\
&=& 
 \int \frac{\mathrm{d} m}{m}\mathrm{e}^{\mathrm{i}\epsilon\ln km} \sum_s \delta\left (\frac{E}{2\pi m}- k\frac{p}{q}- s \right )\mathrm{e}^{2\pi \mathrm{i}sr/k}\nonumber\\
&\approx& \mathrm{e}^{\mathrm{i}\epsilon\ln (Ek/2\pi)} \sum_s \left (\frac{q}{kp-qs}\right )^{1+\mathrm{i}\epsilon}\mathrm{e}^{2\pi \mathrm{i}sr/k}.
\end{eqnarray}
It leads to the following expression for the two-point function
\begin{equation}
R_2^{\,\mathrm{(off)}}(\epsilon)= 
\frac{S(k)}{4\pi^2 k} \mathrm{e}^{\mathrm{i}\epsilon\ln (Ek/2\pi)}\sum_{r=0}^{k-1} g(r) \sum_{(q,k)=1}\sum_{\overset{\scriptstyle{s}}{(p,q)=1}} \left (\frac{q}{kp-qs}\right )^{1+\mathrm{i}\epsilon} 
\mathrm{e}^{2\pi \mathrm{i}s r/k}\left (\frac{\mu(q)}{\phi(q)}\right )^2+\mathrm{c.c.}\; .
\end{equation}
where
\begin{equation}
g(r)=\sum_{r_1=0}^{k-1}\chi(r_1)\bar{\chi}(r_1+r).
\end{equation}
Here we will need the condition that the character $\chi$ is assumed to be primitive. Consider the Gauss sum for a given character $\chi$
\begin{equation}
\tau(\chi)=\sum_{m=1}^k\chi(m)\mathrm{e}^{2\pi \mathrm{i}m/k}
\end{equation}
It is known that for primitive characters the following is true for all $s$ (see e.g. \cite{davenport} sec. 9) 
\begin{equation}
\bar{\chi}(s)\tau(\chi)=\sum_{m=1}^k\chi(m)\mathrm{e}^{2\pi \mathrm{i}ms/k}
\end{equation}
and 
\begin{equation}
|\tau(\chi)|^2=k.
\end{equation}
Using these formulas we get
\begin{equation}
\sum_{r=0}^{k-1} g(r)\mathrm{e}^{2\pi \mathrm{i} rs/k}=
\sum_{r_1,r_2=0}^{k-1}\chi(r_1)\bar{\chi}(r_2)\mathrm{e}^{2\pi \mathrm{i}(r_2-r_1)s /k}=
|\chi(s)|^2 |\tau(\chi)|^2=\left \{\begin{array}{cl} k,&\;(k,s)=1\\0,&\;(k,s)>1
\end{array}\right . .
\end{equation}
Therefore, $s$ has to be coprime to $k$, $(s,k)=1$ and the integer $R=kp-ks$ will be coprime to both $q$ and $k$. As $(q,k)=1$ these restrictions can be written in the form: $(R,qk)=1$.

Finally, the two-point function for the Dirichlet $L$-function with a primitive character is
\begin{equation}
R_2^{\,\mathrm{(off)}}(\epsilon)= 
\frac{S(k)}{4\pi^2} \mathrm{e}^{\mathrm{i}\epsilon\ln (Ek/2\pi)} 
\sum_{(q,k)=1}\sum_{(R,qk)=1}
 \left (\frac{q}{R}\right )^{1+\mathrm{i}\epsilon} 
\left (\frac{\mu(q)}{\phi(q)}\right )^2+\mathrm{c.c.}\; .
\label{off}
\end{equation}
The summation over all integers $R$ coprime with $qk$ is performed by the inclusion--exclusion principle
\begin{equation}
\sum_{(R,qk)=1}f(R)=\sum_{t=1}^{\infty}\sum_{\delta |qk}f(t\delta)\mu(\delta).
\end{equation}
This gives
\begin{equation}
\sum_{(R,qk)=1}\frac{1}{R^{1+\mathrm{i}\epsilon}}=\zeta(1+\mathrm{i}\epsilon)\prod_{p\,|\,q}\left (1-\frac{1}{p^{1+\mathrm{i}\epsilon}}\right )\prod_{p\,|\,k}\left (1-\frac{1}{p^{1+\mathrm{i}\epsilon}}\right ).
\label{R_qk}
\end{equation}
The remaining sum over $q$ coprime to $k$ is calculated as follows 
\begin{eqnarray}
& &\sum_{(q,k)=1} q^{1+\mathrm{i}\epsilon}\left (\frac{\mu(q)}{\phi(q)}\right )^2\prod_{p\,|\,q}\left (1-\frac{1}{p^{1+\mathrm{i}\epsilon}}\right )\nonumber\\
&=&
\prod_{(p,k)=1}\left (1+\frac{1}{p^{1-\mathrm{i}\epsilon}}\frac{1}{(1-1/p)^2}\left (1-\frac{1}{p^{1+\mathrm{i}\epsilon}}\right ) \right )\nonumber\\
&=&\prod_{(p,k)=1}\frac{1}{1-p^{-1+\mathrm{i}\epsilon}}\left (1-\frac{(p^{\mathrm{i}\epsilon}-1)^2}{(p-1)^2} \right )\nonumber\\
&=&\zeta(1-\mathrm{i}\epsilon)\Phi^{\,(\mathrm{off})}(\epsilon)\prod_{p\,|\,k} 
\left ( \frac{1-p^{-1+\mathrm{i}\epsilon}}{1-\frac{(p^{\mathrm{i}\epsilon}-1)^2}{(p-1)^2}}\right )\nonumber\\
&=&\zeta(1-\mathrm{i}\epsilon)\Phi^{\,(\mathrm{off})}(\epsilon)\prod_{p\,|\,k} 
 \frac{p-1}{p(1+(p^{\mathrm{i}\epsilon}-1)/(p-1) )}.
\end{eqnarray}
Here $\Phi^{\,(\mathrm{off})}(\epsilon)$ is the same function as in the Riemann case (\ref{psi_off_riemann}).

Taking into account the expression (\ref{S_k})  for $S(k)$, we find that the total off-diagonal contribution  to the two-point correlation function is
\begin{equation}
R_2^{\,\mathrm{(off)}}(\epsilon)= 
\frac{1}{4\pi^2 } \mathrm{e}^{\mathrm{i}\epsilon\ln (Ek/2\pi)} |\zeta(1+\mathrm{i}\epsilon)|^2
\Phi^{\,(\mathrm{off})}(\epsilon)\Psi^{\,(\mathrm{off})}(\epsilon,k)
+\mathrm{c.c.}\; ,
\label{R_off_dirichlet}
\end{equation}
where
\begin{equation}
\Psi^{\,(\mathrm{off})}(\epsilon,k)=\prod_{p\,|\,k} 
\left (1+\frac{(p^{\mathrm{i}\epsilon/2}-p^{-\mathrm{i}\epsilon/2})^2}{p-p^{-\mathrm{i}\epsilon}}\right )^{-1}.
\label{psi_off_dirichlet}
\end{equation}
Notice that $\Psi^{\,(\mathrm{off})}(\epsilon)\stackrel{\epsilon\to 0}{\longrightarrow} 1$ so the universal limit of the two-point correlation function coincides with the GUE/CUE result.

\section{Conclusion}\label{conclusion}

The two-point correlation function for the Dirichlet $L$-function with a primitive character is as follows
\begin{equation}
R_2(\epsilon)=\bar{d}^2(E)+R_2^{\,\mathrm{(diag)}}(\epsilon)+R_2^{\,\mathrm{(off)}}(\epsilon).
\end{equation}
Here $\bar{d}(E)$ is the mean density of zeros
\begin{equation}
\bar{d}(E)=\frac{1}{2\pi}\ln \left (\frac{kE}{2\pi} \right ).
\end{equation}
The diagonal part of the correlation function, $R_2^{\,\mathrm{(diag)}}(\epsilon)$, and  the off-diagonal part, $R_2^{\,\mathrm{(off)}}(\epsilon)$, have the form \begin{equation}
R_2^{\mathrm{\,(diag)}}(\epsilon)=
-\frac{1}{4\pi^2}\frac{\partial^2}{\partial \epsilon^2}\ln \left [|\zeta(1+\mathrm{i}\epsilon)|^2
\Phi^{\mathrm{\,(diag)}}(\epsilon) \Psi^{\mathrm{\,(diag)}}(\epsilon,k) \right ]
\end{equation}
and
\begin{equation}
R_2^{\,\mathrm{(off)}}(\epsilon)= 
\frac{1}{4\pi^2 } \mathrm{e}^{2\pi \mathrm{i}\epsilon \bar{d}(E)} |\zeta(1+\mathrm{i}\epsilon)|^2
\Phi^{\,(\mathrm{off})}(\epsilon) \Psi^{\,(\mathrm{off})}(\epsilon,k)
+\mathrm{c.c.}\; ,
\end{equation}
where the functions $\Phi^{\mathrm{\,(diag)}}(\epsilon)$ and $\Phi^{\,(\mathrm{off})}(\epsilon)$ are the same as for the Riemann case (cf. (\ref{R_diag_riemann}) and (\ref{R_off_riemann})). The difference is in the functions  $\Psi^{\mathrm{\,(diag)}}(\epsilon,k)$  and  $\Psi^{\,(\mathrm{off})}(\epsilon,k)$  given in (\ref{psi_diag_dirichlet}) and    (\ref{psi_off_dirichlet}) as a finite product of  primes which divide the modulus $k$ of the primitive character used to construct the $L$-function. As all these additional factors tend to 1 when $\epsilon\to 0$ the universal behaviour of the two-point function for the Dirichlet zeros is the same as for the GUE/CUE.

\section{Acknowledgements}
JPK was supported by a grant from the Leverhulme Trust and by the Air Force Office of Scientific Research, Air Force Material Command, USAF, under grant number FA8655-10-1-3088. The U.S. Government is authorized to reproduce and distribute reprints for Governmental purpose notwithstanding any copyright notation thereon.    
 
\appendix
\section{Contribution of the $r=0$ term}\label{zero_term}

For completeness let us consider separately  the term with $t=0$ in (\ref{off}). Then
\begin{equation}
g(0)=\sum_{t_1=0}^{k-1}\chi(t_1)\bar{\chi}(t_1)=\phi(k)
\end{equation} 
and (\ref{off}) takes the form
\begin{equation}
R_2^{\,\mathrm{(off)}}(\epsilon)= 
\frac{S(k)\phi(k)}{4\pi^2 k} \mathrm{e}^{\mathrm{i}\epsilon\ln (Ek/2\pi)} \sum_{(q,k)=1}\sum_{(R,q)=1}\left (\frac{q}{R}\right )^{1+\mathrm{i}\epsilon} +\mathrm{c.c.}\; .
\end{equation}
This differs from the Riemann case by the factor
\begin{equation}
\frac{S(k)\phi(k)}{k}=\frac{k}{\phi(k)}=\prod_{p\,|\,k}\frac{1}{1-p^{-1}},
\end{equation}
and  by the restriction that the summation is performed not over all $q$ but only over $q$ coprime with $k$.

As in (\ref{R_qk}) we get
\begin{equation}
\sum_{(R,q)=1}\frac{1}{R^{1+\mathrm{i}\epsilon}}=\zeta(1+\mathrm{i}\epsilon)\prod_{p\,|\,q}\left (1-\frac{1}{p^{1+\mathrm{i}\epsilon}}\right )
\end{equation}
and 
\begin{eqnarray}
& &\sum_{(q,k)=1} q^{1+\mathrm{i}\epsilon}\left (\frac{\mu(q)}{\phi(q)}\right )^2\prod_{p\,|\,q}\left (1-\frac{1}{p^{1+\mathrm{i}\epsilon}}\right )\nonumber\\
&=&
\prod_{(p,k)=1}\left (1+\frac{1}{p^{1-\mathrm{i}\epsilon}}\frac{1}{(1-1/p)^2}\left (1-\frac{1}{p^{1+\mathrm{i}\epsilon}}\right ) \right )\nonumber\\
&=&\prod_{(p,k)=1}\frac{1}{1-p^{-1+\mathrm{i}\epsilon}}\left (1-\frac{(p^{\mathrm{i}\epsilon}-1)^2}{(p-1)^2} \right )
=\zeta(1-\mathrm{i}\epsilon)\Phi^{\,(\mathrm{off})}(\epsilon)\prod_{p\,|\,k} \left (1-p^{-1+\mathrm{i}\epsilon}\right ).
\end{eqnarray}
Finally we obtain that the contribution of the term with $r=0$ is 
\begin{equation}
R_{2,0}^{\,\mathrm{(off)}}(\epsilon)= 
\frac{1}{4\pi^2 } \mathrm{e}^{\mathrm{i}\epsilon\ln (Ek/2\pi)} |\zeta(1+\mathrm{i}\epsilon)|^2
\Phi^{\,(\mathrm{off})}(\epsilon)\Psi^{\,(\mathrm{off})}(\epsilon,k) \prod_{p\,|\,k} 
\frac{p-p^{-\mathrm{i}\epsilon}}{p-1}
+\mathrm{c.c.}\; .
\end{equation}
This differs from (\ref{R_off_dirichlet}) only by the factor $\prod_{p\,|\,k} 
(p-p^{-\mathrm{i}\epsilon})/(p-1)$ which tends to one when $\epsilon\to 0$ and, therefore, 
$R_{2,0}^{\,\mathrm{(off)}}(\epsilon)$
has exactly the correct singularity when $\epsilon\to 0$. It means that the term with $r=0$ is responsible for the random matrix limit exactly as was shown in \cite{BL} by another method.


\begin{thebibliography}{99}
\bibitem{Mont} H. Montgomery \textit{The pair correlation of the zeta function} in: {\it Proc. Sympos. Pure Math. vol. XXIV, St. Lois. Mo. 1972} Amer. Math. Soc., Providence, R.I. (1973)
\bibitem{Odlyzko} A.M. Odlyzko, \textit{The $10^{20th}$ zero of the Riemann zeta function and 70 million of its neighbours}, Preprint 1989, unpublished
\bibitem{KS} J.P. Keating and N.C. Snaith, \textit{Random matrices and ${L}$-functions}, J. Phys. A: Math. Gen., \textbf{36}, 2859, 2003
\bibitem{HL} G.H. Hardy and J.E. Littlewood, \textit{Some Problems of 'Partitio Numerorum.  III. On the Expression of a Number as a Sum of Primes}, Acta Math. \textbf{44}, 1, 1923
\bibitem{keating_2} J.P. Keating, \textit{Quantum chaology and the Riemann zeta-function}, in Quantum Chaos, eds. G. Casati, I. Guarneri, and U. Smilansky, (North-Holland, Amsterdam), 145, 1993
\bibitem{BK} E.B Bogomolny and  J.P. Keating,  \textit{Random matrix theory and the Riemann zeros I: three- and four-point correlations}, Nonlinearity \textbf{8}, 1115, 1995; ibid \textit{Random matrix theory and the Riemann zeros II: n-point correlations}, Nonlinearity \textbf{9}, 911, 1996
\bibitem{keating} E. B. Bogomolny and J. P. Keating, \textit{Gutzwiller's Trace Formula and Spectral Statistics: Beyond the Diagonal Approximation}, Phys. Rev. Lett. \textbf{77}, 1472, 1996
\bibitem{varenna} E. Bogomolny, \textit{Spectral statistics and periodic orbits}, in Proc. Inter. School of Physics "Enrico Fermi", Varenna, 333,  1999
\bibitem{SIAM} M.V. Berry and J.P. Keating, \textit{The Riemann zeros and eigenvalue asymptotics}, SIAM Review, \textbf{41}, 236, 1999.
\bibitem{osaka} E. Bogomolny, \textit{Riemann zeta function and quantum chaos}, in Proc. Inter. Conference on Quantum Mechanics and Chaos, Osaka Univ. 2006,  Prog. Theor. Phys. Suppl. \textbf{166}, 19, 2007
\bibitem{BL} E. Bogomolny and P. Leboeuf, \textit{Statistical properties of the zeros of zeta functions--beyond the Riemann case}, Nonlinearity \textbf{7}, 1155, 1994
\bibitem{CFKRS} J.B. Conrey, D.W. Farmer, J.P. Keating, M.O. Rubinstein, and N.C. Snaith, \textit{Integral moments of ${L}$-functions}, Proc. London Math. Soc., \textbf{91}, 33, 2005.
\bibitem{CFZ} J.B. Conrey, D.W. Farmer, and M.R. Zirnbauer, \textit{Autocorrelation of ratios of ${L}$-functions}, Comm. Number Theory and Physics, \textbf{2}, 593, 2008.
\bibitem{CS} J.B. Conrey and N.C. Snaith, \textit{Applications of the {$L$}-functions ratios conjectures}, Proc. London Math. Soc., \textbf{94}, 594, 2007.
\bibitem{Hardy} G. H. Hardy, \textit{Note on Ramanujan's trogonometric function $c_q(n)$ and certain series of arithmetical functions}, Proc. Camb. Phil. Soc. \textbf{20}, 263, 1921
\bibitem{davenport} H. Davenport, \textit{Multiplicative number theory}, Springer, 3rd ed., 2000 
\end{thebibliography}
\end{document}